\documentclass{mem}
\usepackage{natbib}\usepackage{txfonts}\usepackage{balance}
\usepackage{graphicx}

\newcommand{\ewli}{${\rm EW}({\rm Li})$}
\newcommand{\teff}{$T_{\rm eff}$}
\newcommand{\msun}{$M_\odot$}

\begin{document}
\def\teff{$T\rm_{eff }$}
\def\kms{$\mathrm {km s}^{-1}$}

\title{
The lithium-rotation connection in young stars
}

   \subtitle{}

\author{
J. \, Bouvier\inst{1} 
          }

\institute{
IPAG, Univ. Grenoble Alpes, 38000 Grenoble,  France\\
\email{Jerome.Bouvier@univ-grenoble-alpes.fr}
}

\authorrunning{Bouvier}

\titlerunning{Lithium-rotation connection}

\abstract{Lithium is a sensitive probe to mixing processes operating in stellar interiors. For many years, a connection has been suspected to exist between lithium abundances and stellar rotation, presumably the result of rotationally-induced internal mixing. In recent years, several studies have confirmed and refined this relationship for low-mass young stars. In various star forming regions and young open clusters, rapidly rotating K dwarfs are found to be lithium-rich compared to their more slowly rotating siblings. While this lithium-rotation correlation is contrary to naive expectations, several models have been put forward to account for it. We review here recent observational results, and briefly discuss proposed interpretations. 
\keywords{Stars: low-mass -- stars: pre-main sequence -- stars: abundances -- stars: rotation -- open clusters and associations}
}
\maketitle{}

\section{Introduction}

Ever since \cite{Skumanich72} reported a parallel decrease of rotation and lithium abundances for solar-type stars on the main sequence, the suspicion has been growing that lithium depletion and rotational evolution are closely linked in low-mass stars. The most enlightening and intriguing result was reported by \cite{Soderblom93} who showed that K-type stars in the young Pleiades open cluster (125~Myr) exhibited a large dispersion of lithium abundances, mimicking their rotational velocity scatter. A tight relationship was suggested where {\it the fastest rotators were the lithium richer stars}. This result was at odds with the predictions of most theoretical models, which would favor strong lithium depletion in fast rotators, as the result of enhanced internal rotational mixing.

As rotation rate measurements of low-mass stars have turned into the industrial era, with tens of thousands of rotational periods being reported thanks to large-scale monitoring campaigns performed from the ground and to the legacy of space missions such as CoRoT and Kepler/K2, the investigation of a connection between lithium and rotation in young stars has known a revival of interest. We report in Section~2 some recent results in this area, and briefly discuss possible scenarios to account for them in Section~3.  

\section{The observational evidence for a lithium-rotation connection in young stars}

\begin{figure*}[t]
\resizebox{\hsize}{!}{\includegraphics[clip=true]{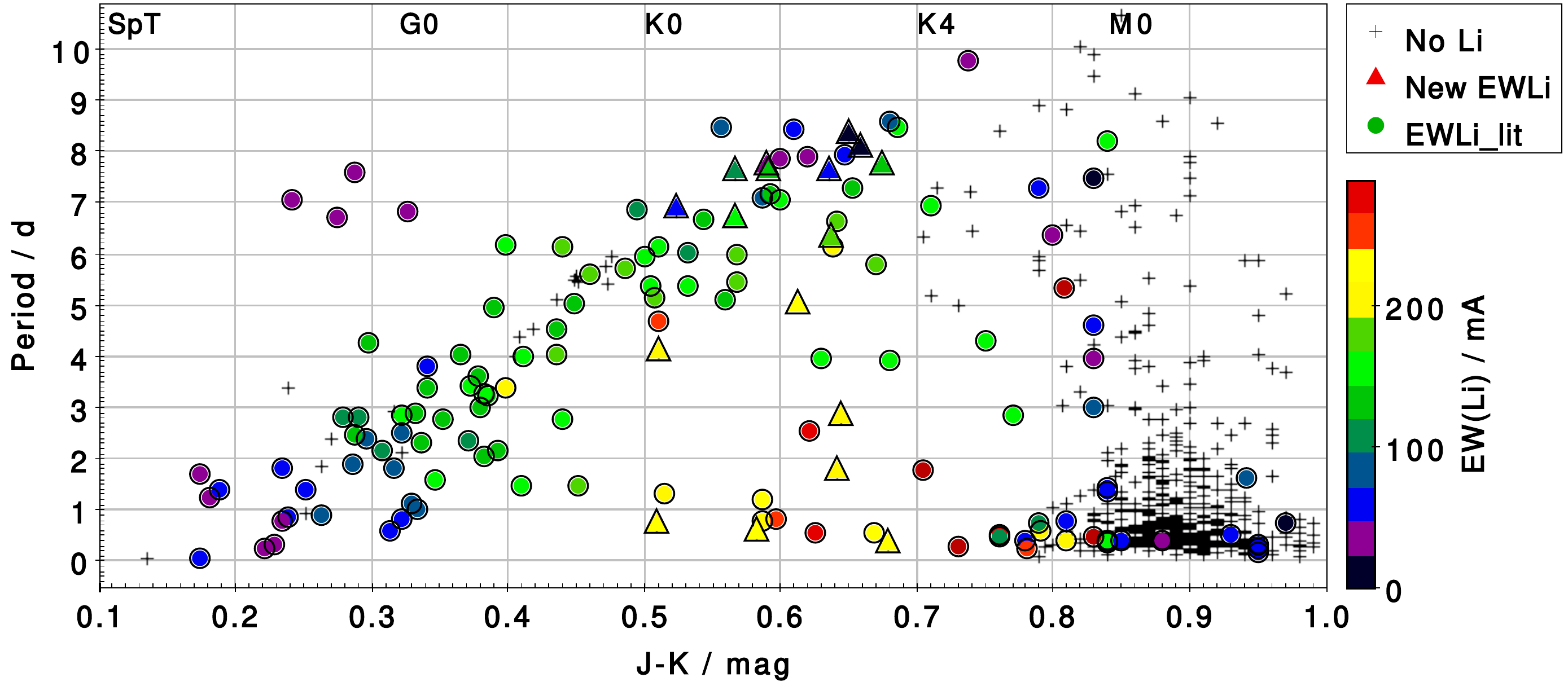}}
\caption{
The distribution of K2 rotational periods from \citet{Rebull16} is shown for members of the $\sim$125~Myr-old Pleiades cluster as a function of (J-K$_s$) color. Crosses are objects without a lithium equivalent width measurement, mostly late-K and M-type stars. Colored symbols indicate lithium equivalent width measurements on a scale which reflects lithium equivalent width (in m\AA), shown on the right side of the plot. Approximate spectral types are indicated at the top of the plot. {\it From \cite{Bouvier18}.} 
}
\label{pleiadesli}
\end{figure*}

In recent years, several star forming regions, moving groups and associations, and young open clusters have been investigated in search for a relationship between lithium content and rotation rate. In all cases, the lithium abundances are derived from the measurements of the equivalent width of the LiI 607.8~nm line, \ewli, which is conspicuous in the spectra of young stars. The rotational periods are derived from high-quality light curves, where surface spots modulate the stellar flux. Both lithium equivalent widths and rotational periods are derived with an accuracy of usually a few percent. 

At the youngest ages of $\sim$5~Myr, as low-mass pre-main sequence stars have just dissipated their circumstellar disks, \cite{Bouvier16} reported a hint of a correlation between \ewli\ and rotation among low-mass members of the NGC~2264 star forming region. Over the \teff\ range from about 3800 to 4400~K ($\sim$0.5-1.2 \msun), the average \ewli\ amounts to about 550-600 m\AA, with a dispersion of about 50~m\AA. While the scatter is relatively modest, a trend emerges for fast rotators to have the largest lithium content at a given \teff. The fastest rotators at this age appear to have preserved their original lithium content, A[Li]$\sim$3.0-3.2~dex, similar to the meteoritic value, while the slower rotators have lithium abundances reduced by about 0.2~dex. 

Slightly further into the pre-main sequence (PMS) stage, \cite{Messina16} reported a correlation between lithium and rotation for low-mass members of the $\sim$24~Myr-old $\beta$ Pic moving group. The average \ewli\ for stars with masses around 0.7~\msun\ has decreased to about 300~m\AA, as models predict moderate PMS lithium burning over this mass range \citep[e.g.][]{Jeffries14}. The dispersion in \ewli\ at a given \teff\ is now about twice as large as what was reported for NGC~2264. A correlation is seen between \ewli\ and rotational period over the mass range from 0.3 to 0.8~\msun, with faster rotators having larger lithium abundances than slower ones, as reported for NGC~2264.  
 
 \begin{figure*}[t]
\resizebox{\hsize}{!}{\includegraphics[clip=true]{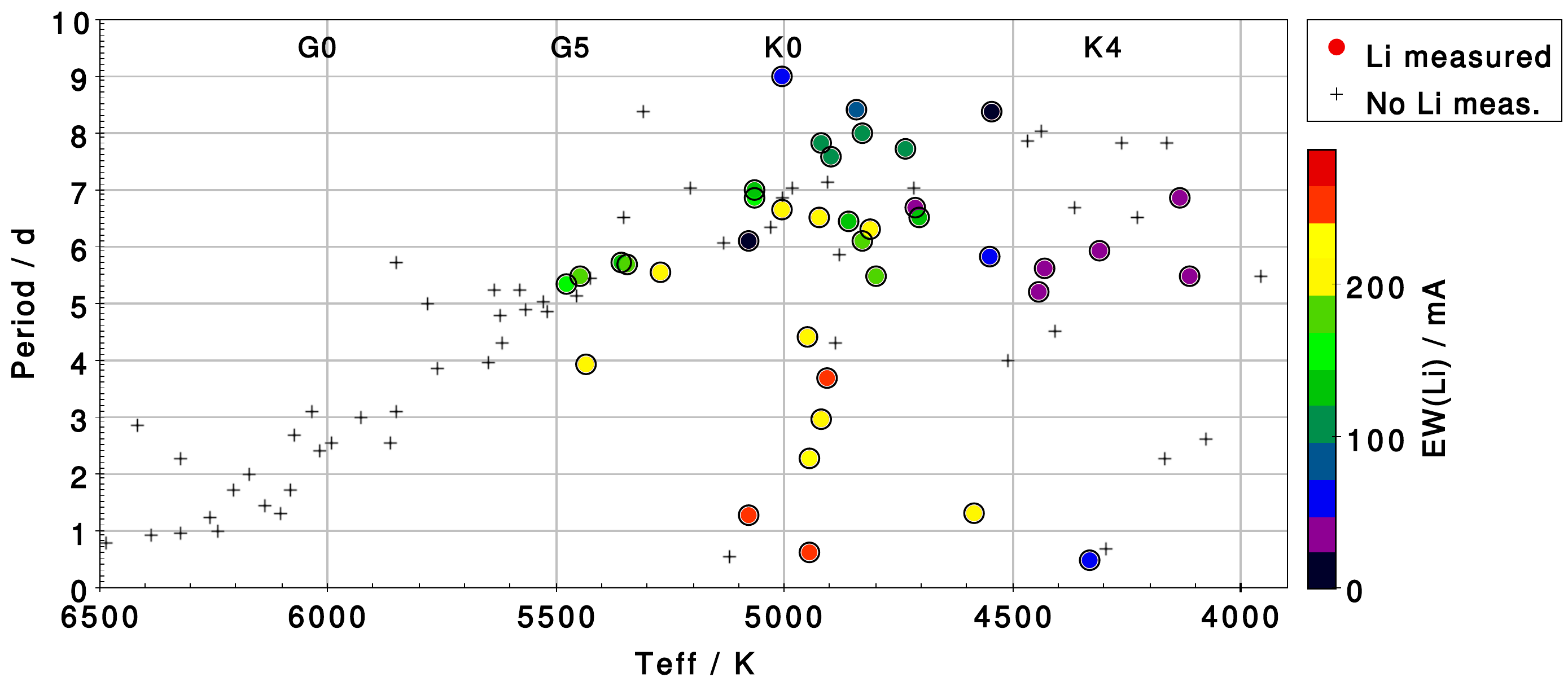}}
\caption{
The distribution of K2 rotational periods from \citet{Curtis19} is shown  as a function of \teff\ for members of the $\sim$125~Myr-old Psc-Eri association.   Crosses are objects without a lithium equivalent width measurement. Circles indicate LiI 670.8 nm measurements with a color that scales with lithium equivalent width (in m\AA), as shown on the right side of the plot. Approximate spectral types are indicated at the top of the plot. {\it From \cite{Arancibia20}.} 
}
\label{pscerili}
\end{figure*}

The most striking relationship between lithium and rotation is that seen in the Pleiades young open cluster, at an age of $\sim$125 Myr. Originally reported by \cite{Soderblom93}, it was recently revisited and completed by \cite{Barrado16} and \cite{Bouvier18}. Figure~\ref{pleiadesli} shows the distribution of rotational periods as a function of spectral type for the Pleiades low-mass members, from late-F to M stars. The rotational periods derived from K2 light curves \citep{Rebull16} have exquisite precision. There is a wide dispersion of rotation rates among early to mid K-type stars, ranging from a fraction of a day to about 10 days, the end result of pre-main sequence angular momentum evolution \citep{Gallet13, Gallet15}. Except for M dwarfs, most stars have lithium measurements. K dwarfs display a large dispersion of \ewli, which mimics that seen for their rotation rates. In the K0-K4 range (0.75-0.90~\msun), all fast rotators with periods less than 4 days have large \ewli, while the slowest rotators, with periods from 6 to 10 days, have low lithium content.  The corresponding abundances range from A[Li]$\sim$2.8 for fast rotators down to A[Li]$\sim$1.2 for slow ones \citep{Bouvier18}. 
 
Lately, \cite{Arancibia20} explored the lithium-rotation relationship in the newly discovered Psc-Eri moving group \citep{Meingast19}. Based on its rotational period distribution, \cite{Curtis19} assigned to this young association the same age as the Pleiades open cluster. The results are shown in Figure~\ref{pscerili}, in the same format as adopted for the Pleiades cluster illustrated in Figure~\ref{pleiadesli}. Even though the measurements are scarcer than for the Pleiades cluster, a similar pattern of rotation-dependent \ewli\ appears for early K-type stars.  
Presumably, the Pleiades cluster formed from a very rich and dense environment, while the Psc-Eri stream likely originated from a low density association at birth. These drastically different initial conditions have apparently had no impact on the lithium-rotation connection observed at the zero-age main sequence. This suggests that {\it the origin of the lithium-rotation relationship is to be found in the physics of pre-main sequence stellar evolution}, and is not imprinted by specific initial conditions and/or environment.

Finally, Jeffries' presentation (this volume) highlights the fate of the lithium-rotation relationship in the young open cluster M35, which is slightly older than the Pleiades cluster, at an age of about 150~Myr. The lithium-rotation connection is still present at this age, and the wealth of rotational and lithium measurements for this cluster may provide additional clues to its origin. The reader is referred to this and other contributions in these proceedings related to PMS lithium evolution.  

\bigskip

\section{Physical scenarios for the origin of the lithium-rotation connection}
 
 The overall picture that emerges from the observational results described above is that a dispersion of lithium abundances appears early-on during pre-main sequence evolution, within a restricted mass domain from about 0.5 to 1.0~\msun, and is closely associated to the rotation rates of young stars, in the sense that higher lithium abundances are measured for faster rotators. Indeed, the development of the lithium scatter during the PMS seems to go hand in hand with the development of a similar dispersion in rotation rates, which culminates at the zero-age main sequence \citep{Gallet13, Gallet15}. Based on these results, several scenarios have been put forward in an attempt to physically link lithium and angular momentum evolution during the pre-main sequence.  
 
 The most straightforward interpretation relies on the direct impact of rotation on stellar structure and internal transport processes. Thus, \cite{Siess97} empirically suggested that convective mixing efficiency could be impacted by rotation. Assuming that the mixing-length parameter $\alpha$ scales with the inverse of the angular velocity, they showed that the fastest rotators would reach the zero-age main sequence with the highest lithium abundances. More recently, based on 2D numerical simulations of convective overshooting, \cite{Baraffe17} suggested the existence of a threshold in stellar rotation rate above which rotation strongly prevents the vertical penetration of convective plumes in the radiative core, a process that would reduce lithium depletion in fast rotators during the pre-main sequence. 
 
A variant of the above scenario considers the impact of rotationally driven magnetic activity. The stronger the magnetic field, the less efficient is convective transport. Hence, enhanced activity has the same effect of reducing the mixing-length parameter, thus leading to inflated stellar radius and a lower temperature of the base of the convective zone, thus reducing lithium depletion \citep{Ventura98, Chabrier07, Jeffries17}. \cite{Somers14, Somers15} computed models along these lines, arguing that if magnetic activity is linked to rotation during the pre-main sequence, fast rotators will undergo significant radius inflation, and a reduced lithium depletion rate. Whether magnetic activity scales with rotation in pre-main sequence stars remains however to be established \citep{Lavail19,Sokal20}
 
As an alternative to internal processes, \cite{bouvier08} investigated the impact of star-disk interaction on PMS lithium abundances. He suggested that stars that remain locked to their disk for a long time, up to a few million years, would develop strong levels of internal differential rotation: while the outer convective zone is braked by the magnetic interaction with the circumstellar disk \citep[e.g.,][]{Gallet19}, the growing radiative core spins up. A long disk lifetime thus simultaneously prevents the young star from spinning up and enhances lithium depletion due to externally enforced rotational mixing. As a result, slow rotators reach the main sequence as lithium-depleted stars compared to their faster counterparts \citep{Eggenberger12b}. 

\section{Conclusions}

Mounting evidence for a tight relationship between lithium content and rotation rate in young stars has been reported in the last years. It is likely that the lithium-rotation connection observed on the zero-age main sequence, and its earlier development during the pre-main sequence, reflects feedback processes between early angular momentum evolution and lithium depletion. How both quantities physically impact each other is still a matter of debate, and improved physical models of PMS evolution including rotational and magnetic effects, as well as external influences, such as disk accretion, are warranted. In parallel, the wealth of rotational data expected from new photometric space missions (e.g., TESS, PLATO) and the development of large-scale multi-object spectrographs (e.g., MSE, 4MOST, WEAVE) open the way to investigate the lithium-rotation connection in a variety of environments over a broad range of stellar ages and metallicities, which will undoubtedly help to understand the physics it conceals.

\begin{acknowledgements}
It is a pleasure to thank the organizers for their kind invitation to this conference. It has been a unique opportunity to enlarge the scope of the lithium problem(s) and discuss it in a variety of contexts. This research has received funding from the European Research Council (ERC) under the European Union's Horizon 2020 research and innovation programme (grant agreements No 742095 SPIDI: Star-Planets-Inner Disk-Interactions, http://spidi-eu.org).
\end{acknowledgements}

\bibliographystyle{aa} 
\bibliography{liroma} 

\end{document}